\documentclass[twocolumn,aip,reprint,numerical]{revtex4-1}
\usepackage[latin9]{inputenc}
\usepackage[a4paper]{geometry}
\usepackage{array}
\usepackage{float}
\usepackage{textcomp}
\usepackage{bm}
\usepackage{tipa}
\usepackage{amsmath}
\usepackage{amssymb}
\usepackage{graphicx}

\makeatletter

\newcommand{\lyxmathsym}[1]{\ifmmode\begingroup\def\b@ld{bold}
  \text{\ifx\math@version\b@ld\bfseries\fi#1}\endgroup\else#1\fi}

\providecommand{\tabularnewline}{\\}

\usepackage{dcolumn}
\usepackage{bm}
\usepackage{epsfig}
\usepackage{braket}

\usepackage{lineno}
\makeatother

\begin{document}

\title{Field-free Three-Dimensional Orientation of Asymmetric-Top Molecules}

\author{Kang Lin}

\affiliation{State Key Laboratory of Precision Spectroscopy, East China Normal
University, Shanghai 200062, China}

\author{Ilia Tutunnikov}

\affiliation{Department of Chemical and Biological Physics, Weizmann Institute
of Science, Rehovot 7610001, Israel}

\author{Junjie Qiang}

\affiliation{State Key Laboratory of Precision Spectroscopy, East China Normal
University, Shanghai 200062, China}

\author{Junyang Ma}

\affiliation{State Key Laboratory of Precision Spectroscopy, East China Normal
University, Shanghai 200062, China}

\author{Qiying Song}

\affiliation{State Key Laboratory of Precision Spectroscopy, East China Normal
University, Shanghai 200062, China}

\author{Qinying Ji}

\affiliation{State Key Laboratory of Precision Spectroscopy, East China Normal
University, Shanghai 200062, China}

\author{Wenbin Zhang}

\affiliation{State Key Laboratory of Precision Spectroscopy, East China Normal
University, Shanghai 200062, China}

\author{Hanxiao Li}

\affiliation{State Key Laboratory of Precision Spectroscopy, East China Normal
University, Shanghai 200062, China}

\author{Fenghao Sun}

\affiliation{State Key Laboratory of Precision Spectroscopy, East China Normal
University, Shanghai 200062, China}

\author{Xiaochun Gong}

\affiliation{State Key Laboratory of Precision Spectroscopy, East China Normal
University, Shanghai 200062, China}

\author{Hui Li}

\affiliation{State Key Laboratory of Precision Spectroscopy, East China Normal
University, Shanghai 200062, China}

\author{Peifen Lu}

\affiliation{State Key Laboratory of Precision Spectroscopy, East China Normal
University, Shanghai 200062, China}

\author{Heping Zeng}

\affiliation{State Key Laboratory of Precision Spectroscopy, East China Normal
University, Shanghai 200062, China}

\author{Yehiam Prior}

\affiliation{Department of Chemical and Biological Physics, Weizmann Institute
of Science, Rehovot 7610001, Israel}

\author{Ilya Sh. Averbukh}

\affiliation{Department of Chemical and Biological Physics, Weizmann Institute
of Science, Rehovot 7610001, Israel}

\author{Jian Wu}

\affiliation{State Key Laboratory of Precision Spectroscopy, East China Normal
University, Shanghai 200062, China}

\affiliation{Collaborative Innovation Center of Extreme Optics, Shanxi University,
Taiyuan, Shanxi 030006, China}

\maketitle
\textbf{Alignment and orientation of molecules by intense, ultrashort
laser fields are crucial for a variety of applications in physics
and chemistry. These include control of high harmonics generation
\cite{Velotta2001,Itatani2005}, molecular orbitals tomography \cite{Itatani2005,Patchkovskii2007,Lein2007},
control of molecular photoionization and dissociation processes \cite{Larsen1999,Tsubouchi2001,Litvinyuk2003},
production of ``molecular movies\textquotedblright{} with the help
of X-ray free-electron laser sources and ultrafast electron diffraction
of relativistic electrons \cite{Kupper2014,Glownia2016,Yang2016}.
While the dynamics of laser-induced molecular alignment has been extensively
studied and demonstrated (for a review, see \cite{StapelfeldtSeidman2003,Ohshima2010,Fleischer2012,Lemeshko2013}),
molecular orientation is a much more challenging task, especially
for asymmetric-top molecules. Here we report the first experimental
demonstration of a field-free, all-optical three-dimensional orientation
of asymmetric-top molecules by means of phase-locked cross-polarized
two-color laser pulses. In addition to the conventional integrated
orientation factor, we report the differential degree of orientation
which is not amenable to optical measurements, but is readily accessible
in our angle-resolved imaging technique. Our scheme applies to a wide
class of asymmetric molecules and opens new ways towards controlling
their orientation, eventually leading to direct imaging of structure
of gas-phase molecules \cite{molimaging} using advanced free electron
laser beams with extremely high spatiotemporal resolution. }

Over the years, several approaches have been used to break the symmetry
and define preferred directions in space so that molecules can be
oriented. Early on, intense nonresonant laser fields were combined
with weak electrostatic fields \cite{Friedrich1999,Sakai2003,Ghafur2009,Holmegaard2009,Goban2008}
for symmetry breaking, and this was followed by introduction of single-cycle
THz pulses \cite{Harde1991,Dion1999,Averbukh2001,Machholm2001,Fleischer2011,Kitano2013},
alone or in combination with optical pulses \cite{Daems2005,Gershnabel2006,Egodapitiya2014,Damari2016}.
More recently, pulsed laser fields with twisted polarization were
shown to be effective for orienting generic asymmetric molecules \cite{Gershnabel2018},
and for enantio-selective orientation of chiral molecules \cite{Yachmenev2016,Tutunnikov2018}.
Of special interest for us is an all-optical approach that uses nonresonant
two-color laser field (fundamental wave and its second harmonic) where
the orientation is achieved via the nonlinear interaction with the
molecular hyperpolarizability \cite{Vrakking1997,Dion1999,Kanai2001,De2009,Oda2010,JW2010,Frumker2012,Takemoto2008}.

In the present work, we report the first experimental demonstration
of all-optical field-free three-dimentional (3D) orientation of asymmetric-top
molecules. We introduce a scheme using phase-locked Orthogonal Two-Color
(OTC) laser fields, which is applicable to a large class of polyatomic
molecules. In our experiments, the fundamental field aligns the major
molecular axis (the one with the highest polarizability) along the
polarization direction. The second harmonic field is temporally overlapping
with the fundamental field. The two fields together couple to the
molecule via off-diagonal components of the molecular hyperpolarizability
tensor. This interaction orients the minor molecular axis (the one
with the second highest polarizability) after the OTC pulse is over
and results in field-free, 3D molecular orientation. In what follows,
we describe the experimental setup, show the results of experimental
observations, provide a simplified two dimensional (2D) model that
illustrates the principles of our method, and discuss the results
of a more sophisticated fully three-dimensional simulation of the
experimental measurements.

\begin{figure}[h]
\begin{centering}
\includegraphics[scale=0.97]{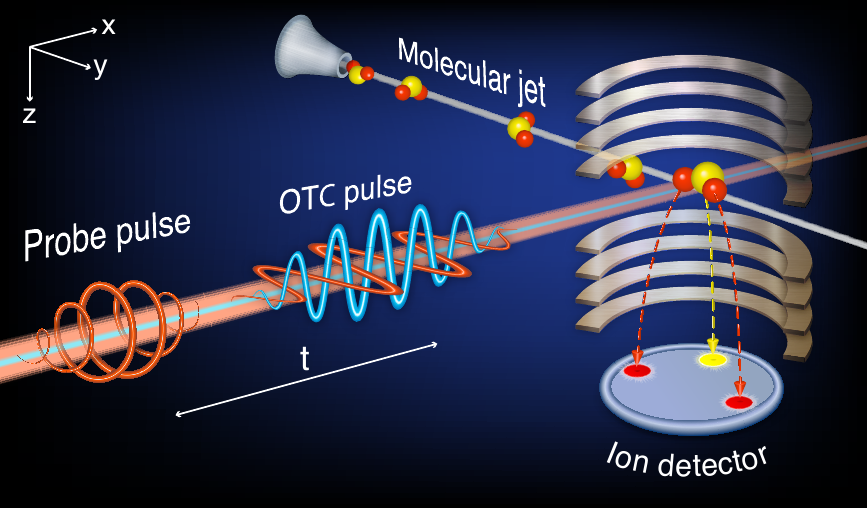}
\par\end{centering}
\caption{\textbf{Layout of the experiment. }A supersonic gas jet of SO$_{2}$
molecules subject to a pair of synchronized two-color laser pulses
with orthogonal polarizations in an ultrahigh vacuum chamber of COLTRIMS.
%By scanning the delay of an intense circularly polarized probe pulse, snapshots of the angular distribution of the molecular axis at various times are taken via Coincident Coulomb explosion.
\label{fig:Fig 1 - Experimental Setup}}
\end{figure}

\begin{figure}[h]
\begin{centering}
\includegraphics[scale=0.09]{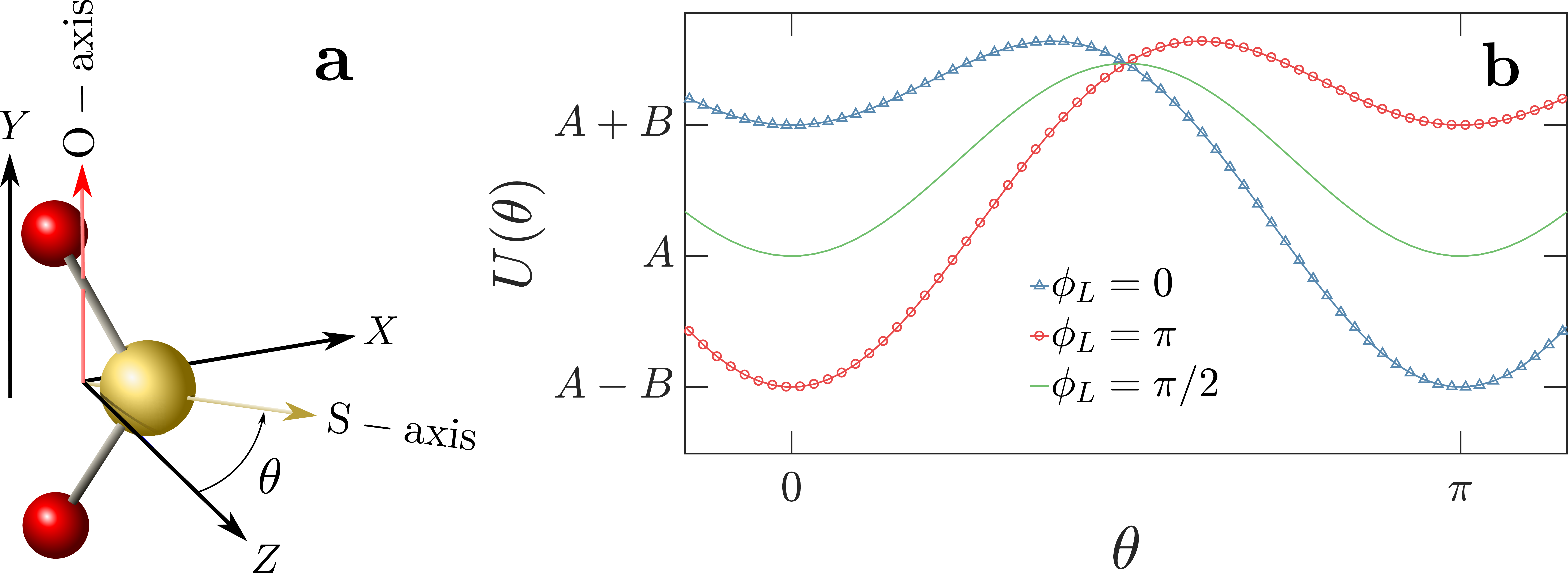}
\par\end{centering}
\caption{\textbf{2D model. a}, $\mathrm{SO}_{2}$ molecule whose major axis
is perfectly aligned along the laboratory Y axis. The minor axis lies
in the $XZ$ plane at an angle $\theta$ relative to the $Z$ axis.
\textbf{b}, Potential energy as a function of $\theta$ for $\phi_{L}=0,\;\pi,\;\pi/2$,
(see eqs. \ref{eq:U-afo-theta-special} and \ref{eq:approximated-potential})
\label{fig:Fig 2 - 1D Model}}
\end{figure}

The experiments were performed on Sulphur dioxide ($\mathrm{SO}_{2}$)
molecules in a supersonic molecular beam (rotational temperature $T=18\;K$).
The O-axis (see Fig. \ref{fig:Fig 2 - 1D Model}a) is the major molecular
axis with the largest polarizability. The S-axis bisects the bond
angle between the oxygen atoms, and it is the minor axis with the
second highest polarizability. The molecular permanent dipole moment
is directed along the S-axis. In our scheme, the laser pulses propagate
along the $X$ axis. The fundamental field is a $Y$-polarized femtosecond
laser pulse with the fundamental wavelength (FW) of 790 nm. The second
color field is a $Z$-polarized second harmonic (SH) pulse of 395
nm whose frequency is doubled relative to FW. It temporally and spatially
overlaps the first pulse, and is phase locked to it. The OTC pulse
is focused on a supersonic molecular beam propagating along the $Y$
axis. At a variable delay after the application of the two-color pulse,
an intense circularly polarized probe pulse is employed to explode
the molecules and to image their 3D spatial orientation via coincident
Coulomb explosion imaging technique, as is schematically shown in
Fig. \ref{fig:Fig 1 - Experimental Setup}. The reaction miscroscope
of Cold Target Recoil Ion Momentum Spectroscopy (COLTRIMS) setup \cite{DORNER200095}
provides direct access to the spatiotemporal molecular dynamics as
a function of the time delay after the OTC pulses with femtosecond
time-resolution. The details of the experimental system are given
in Methods Subsection \ref{subsec:methods-1}.

To illustrate the two-color orientation mechanism, we consider a classical
ensemble of cold $\mathrm{SO}_{2}$ molecules whose major axis is
perfectly aligned along the laboratory $Y$ axis with a uniform angular
distribution of the molecular minor axis in the $XZ$ plane. The angle
of rotation around the alignment axis is denoted by $\theta$, see
Fig. \ref{fig:Fig 2 - 1D Model}a. At time $t=0$, a short phase-locked
OTC pulse is applied to the ensemble. The electric field is given
by
\begin{equation}
\boldsymbol{\mathcal{E}}=\mathcal{E}_{1}(t)\cos(\omega t)\mathbf{e}_{Y}+\mathcal{E}_{2}(t)\cos(2\omega t+\phi_{L})\mathbf{e}_{Z},\label{eq:two-color-laser}
\end{equation}
where $\mathbf{e}_{Y,Z}$ are the units vectors along the corresponding
laboratory axes, $\mathcal{E}_{i}(t)$ are fields' envelopes, $\omega$
is the carrier frequency of the FW field and $\phi_{L}$ is the relative
phase between the FW and SH fields. The potential, $U$ describing
the interaction of the laser field with the molecules (to third order
in the electric field) is given by \cite{MolecularHyperpolarizabilities,SO2-Data}
\begin{equation}
U=-\frac{1}{2}\alpha_{ij}E_{i}E_{j}-\frac{1}{6}\beta_{ijk}E_{i}E_{j}E_{k},\label{eq:interaction-potential}
\end{equation}
where $\alpha_{ij}$ are the components of the polarizability tensor,
$\beta_{ijk}$ are the hyperpolarizability tensor components, and
$E_{i}$ are the components of the electric field. The tensors $\alpha_{ij}$
and $\beta_{ijk}$ are symmetric in all indices \cite{MolecularHyperpolarizabilities}
and summation over the repeated indices is implied. For molecules
having $C_{2v}$ symmetry the hyperpolarizability tensor has three
independent elements $\beta_{113}$, $\beta_{223}$ and $\beta_{333}$.
For $\mathrm{SO}_{2}$ molecule, index $1$ corresponds to the major
O-axis, $3$ to the minor S-axis and $2$ to third principal axis
that points out of the molecular plane \cite{MolecularHyperpolarizabilities,SO2-Data}.

Explicit expression for the interaction energy averaged over fast
optical oscillations as a function of the angle $\theta$ is given
by:
\begin{equation}
U\left(\theta\right)=-\mathcal{E}_{20}^{2}a\cos(2\theta)+\mathcal{E}_{10}^{2}\mathcal{E}_{20}b\cos(\theta),\label{eq:U-afo-theta-special}
\end{equation}
where $a=\left(\alpha_{33}-\alpha_{22}\right)/8$, $b=\beta_{113}\cos(\phi_{L})/8$
and $\mathcal{E}_{i0}$ are the amplitudes of the fields. For $\mathrm{SO}_{2}$
molecules, $\alpha_{22}<\alpha_{33}$, and $\beta_{113}>0$. The $\cos(2\theta)$
term arises from the field interaction with the linear molecular polarizability,
while the $\cos\left(\theta\right)$ term results from the hyperpolarizability
interaction. Figure \ref{fig:Fig 2 - 1D Model}b shows $U\left(\theta\right)$
for various $\phi_{L}$ values. As seen, the potential is a tilted
double well with the tilt controlled by the relative phase $\phi_{L}$.
When the relative phase is $\phi_{L}=\pi/2$, then $b=0$ and the
graph of the potential is the solid-green curve (Fig. \ref{fig:Fig 2 - 1D Model}b).
In this case, the potential is a symmetric function of $\theta$ in
the interval $\left[0,\pi\right]$ and its two minima are equivalent.
A kick by such a potential leads the focusing of the angular distribution
at $\theta=0,\pi$ shortly after the kick.\cite{Averbukh2001} If
$\phi_{L}\neq\pi/2$, the symmetry of the potential function is broken
and the minima are no longer equivalent, manifested in asymmetrical
evolution of the angular distribution. Near the minima ($\theta=0,\pi$),
the potential may be approximated as
\begin{equation}
U\left(\theta\right)\approx\begin{cases}
\left(-A+B\right)+\kappa_{-}\theta^{2} & \theta\approx0\\
\left(-A-B\right)+\kappa_{+}\left(\theta-\pi\right)^{2} & \theta\approx\pi
\end{cases},\label{eq:approximated-potential}
\end{equation}
where $A=\mathcal{E}_{20}^{2}a$, $B=\mathcal{E}_{10}^{2}\mathcal{E}_{20}b$
and $\kappa_{\mp}=\left[4A\mp B\right]/2$.

\begin{figure}[h]
\includegraphics[scale=0.095]{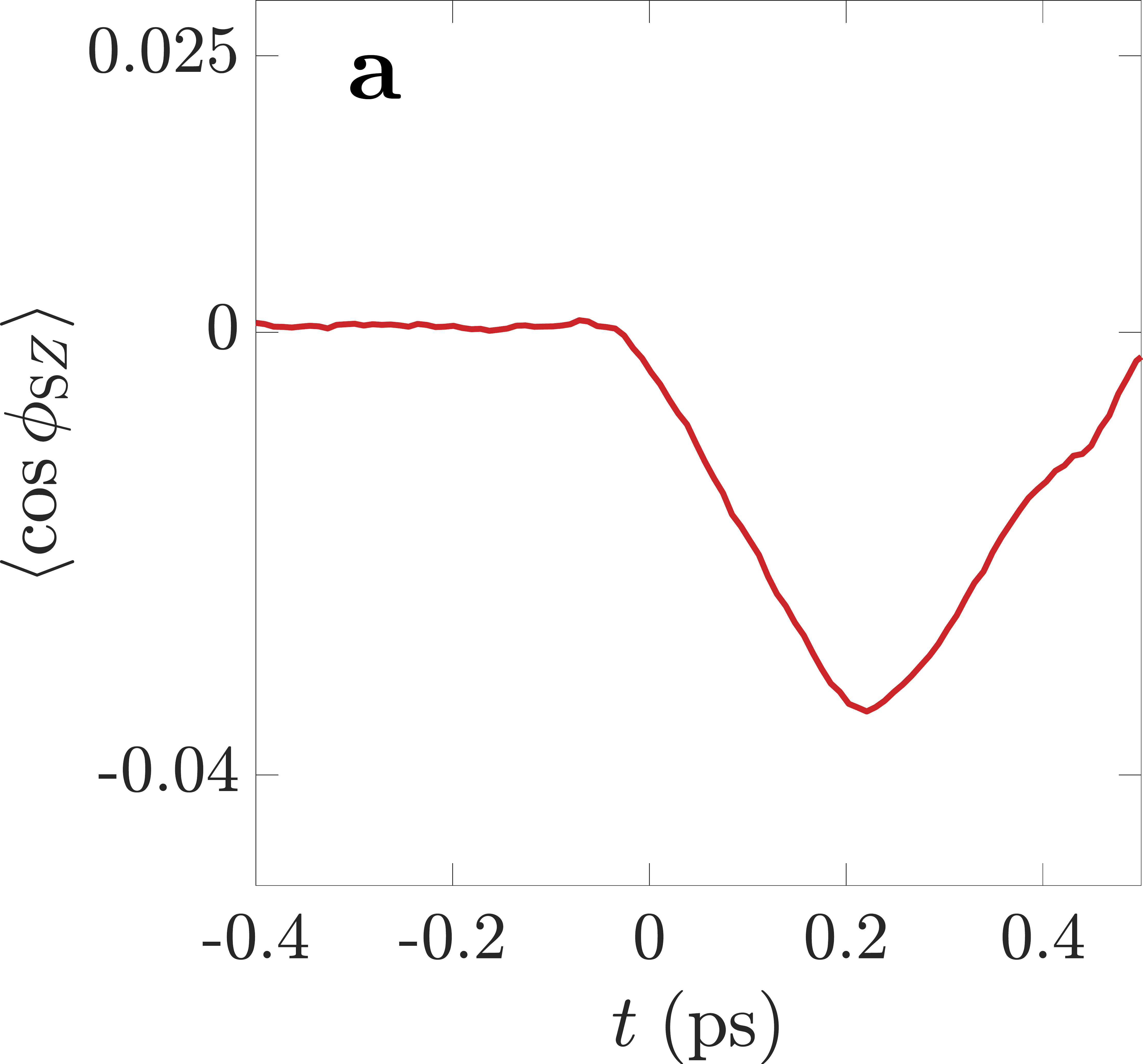}$\;$\includegraphics[scale=0.095]{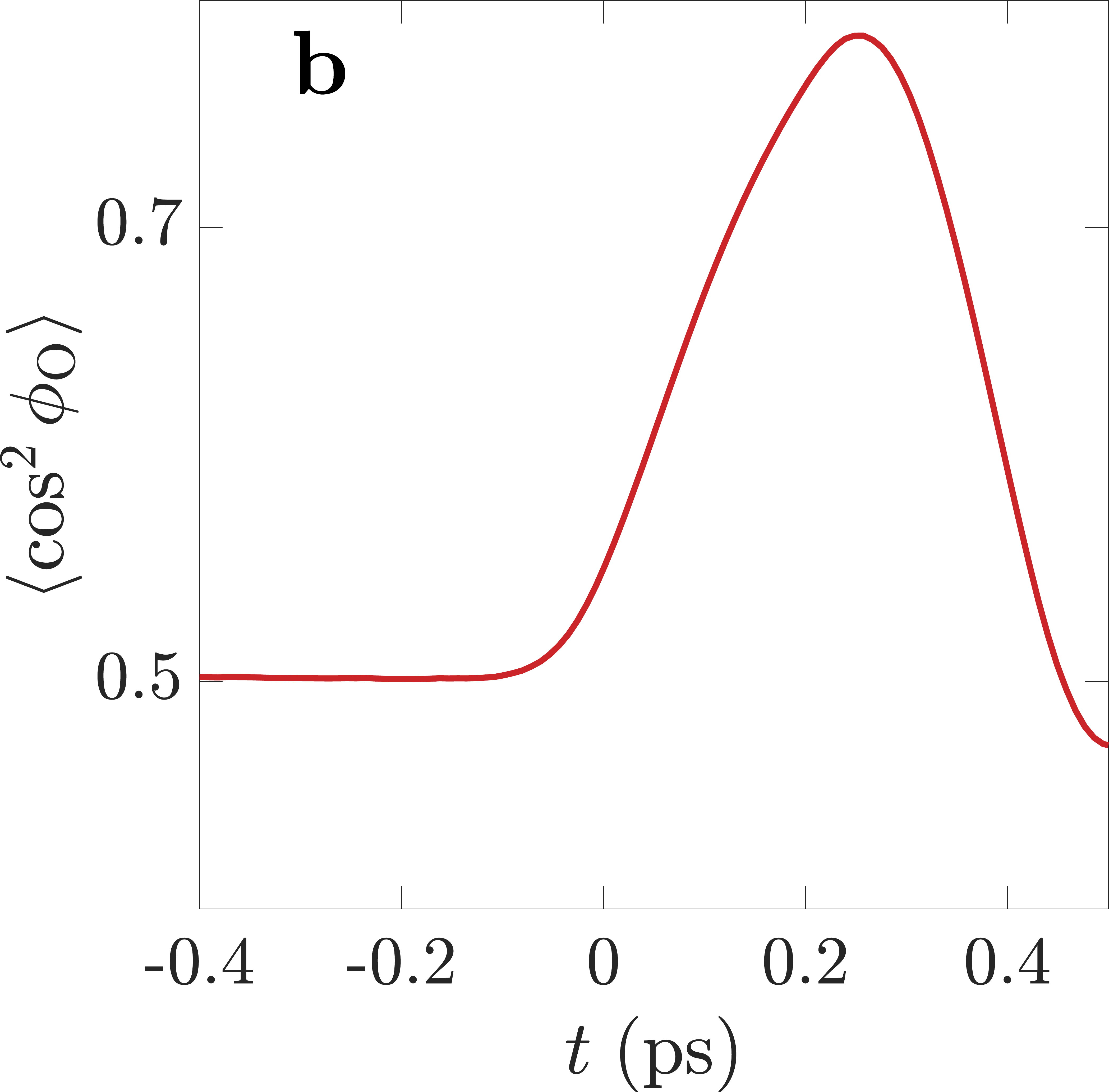}
\begin{raggedleft}
\includegraphics[scale=0.12]{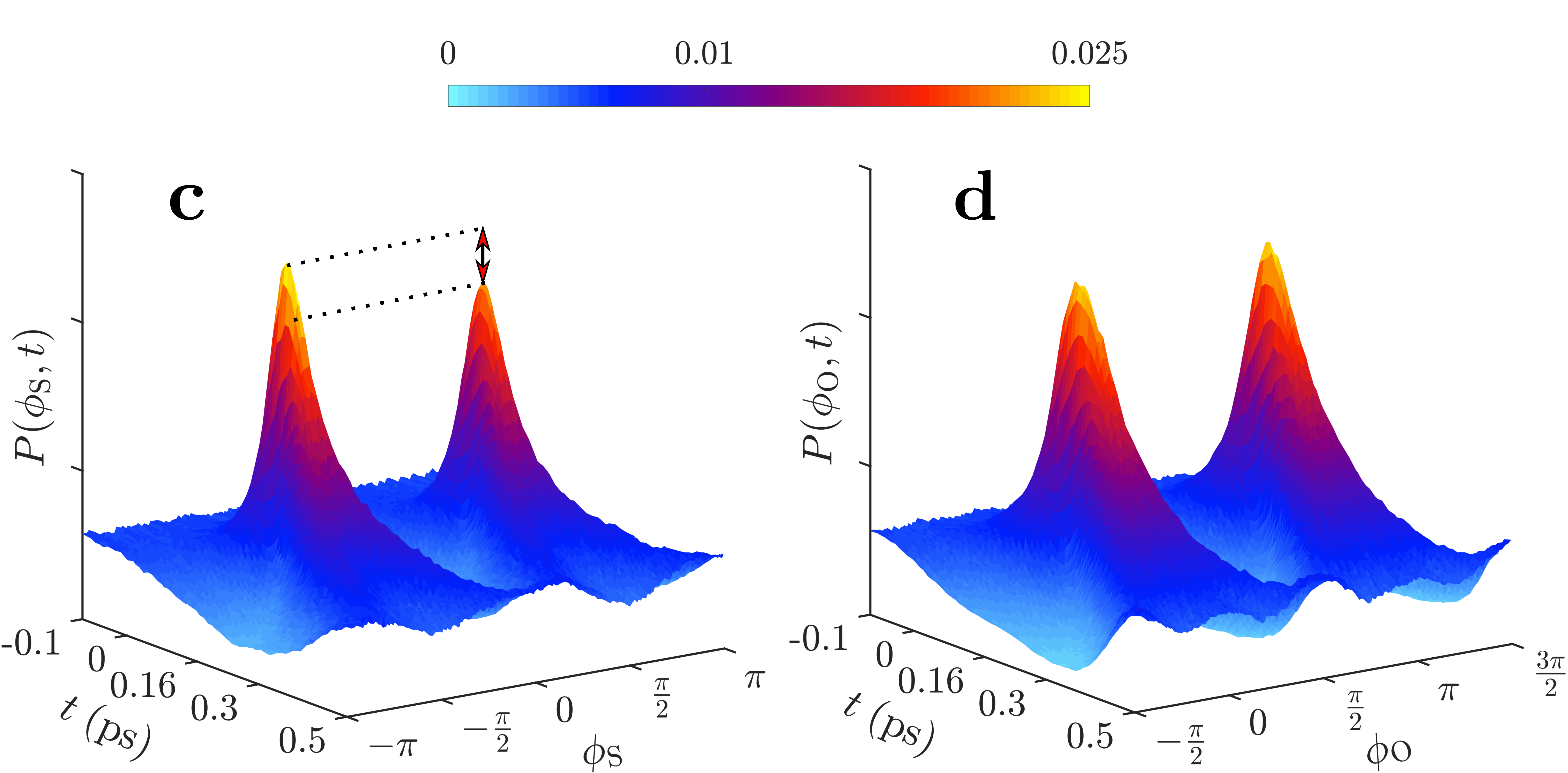}
\par\end{raggedleft}
\begin{raggedright}
\includegraphics[scale=0.165]{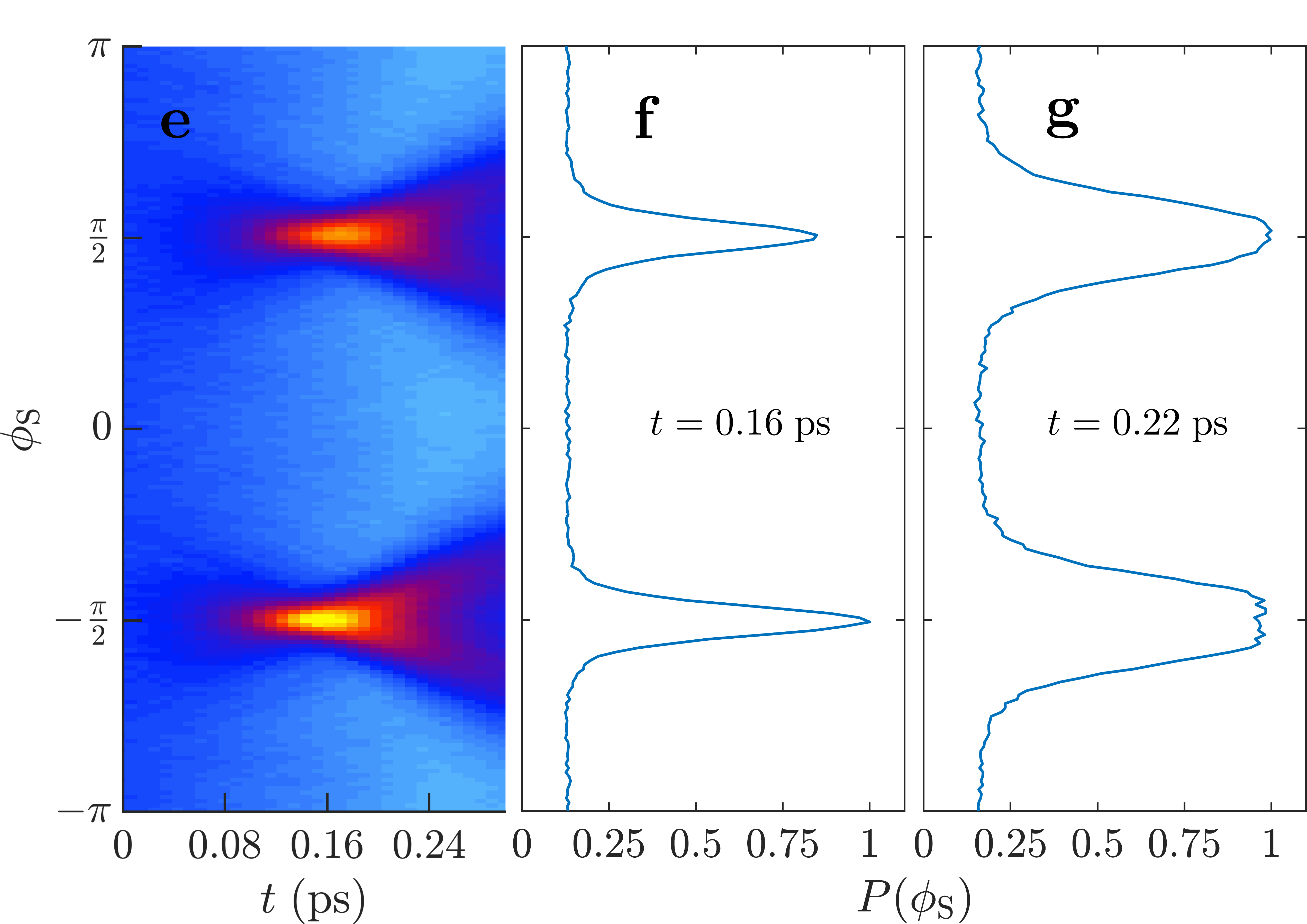}
\par\end{raggedright}
\caption{\textbf{Results of 3D simulation.} \textbf{a}, Orientation factor
$\langle\cos\phi_{\mathrm{S}Z}\rangle\left(t\right)$ \textbf{b},
Alignment factor $\langle\cos^{2}\phi_{\mathrm{O}}\rangle\left(t\right)$.
\textbf{c}, 3D surface plot of the time dependent angular distribution
for S-axis. \textbf{d}, 3D surface plot of the time dependent angular
distribution for O-axis. \textbf{e}, Top view of the 3D surface \textbf{c}.
\textbf{f}, Angular distribution, $P\left(\phi_{\mathrm{S}}\right)$
at $t=0.16\;\mathrm{ps}$, the moment of tight focus at $\phi_{\mathrm{S}}=-\pi/2$.
\textbf{g}, Angular distribution of, $P\left(\phi_{\mathrm{S}}\right)$
at $t=0.22\;\mathrm{ps}$, the moment of maximal orientation. \label{fig:Figure 3 - Results of 3D Simulation}}
\end{figure}
\begin{figure*}[t]
\begin{centering}
\includegraphics[scale=0.36]{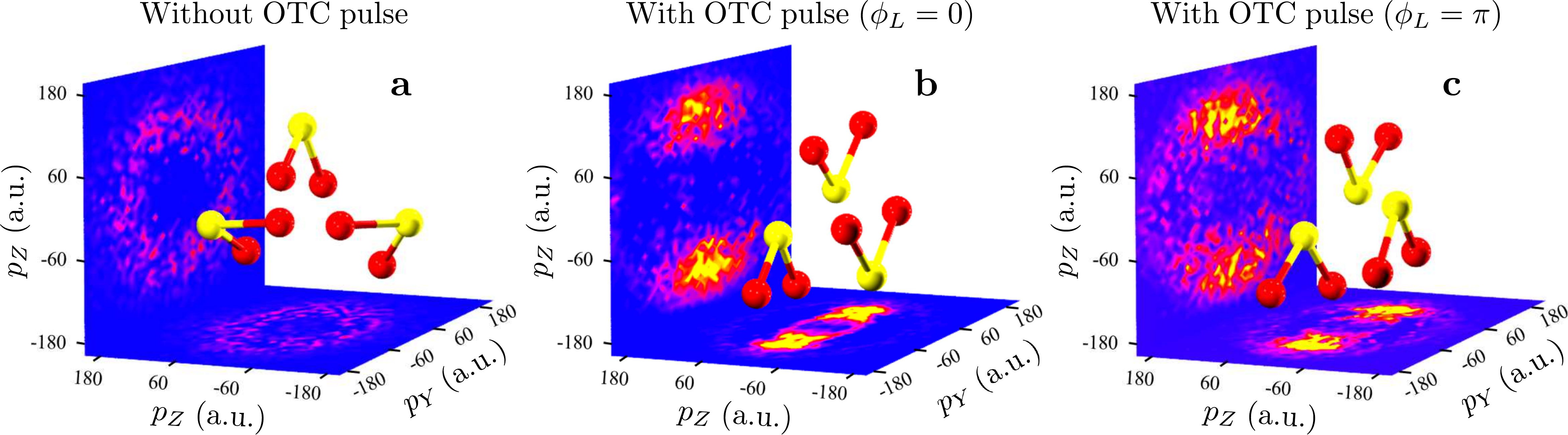}
\par\end{centering}
\begin{centering}
$\;$
\par\end{centering}
\begin{centering}
\includegraphics[scale=0.4]{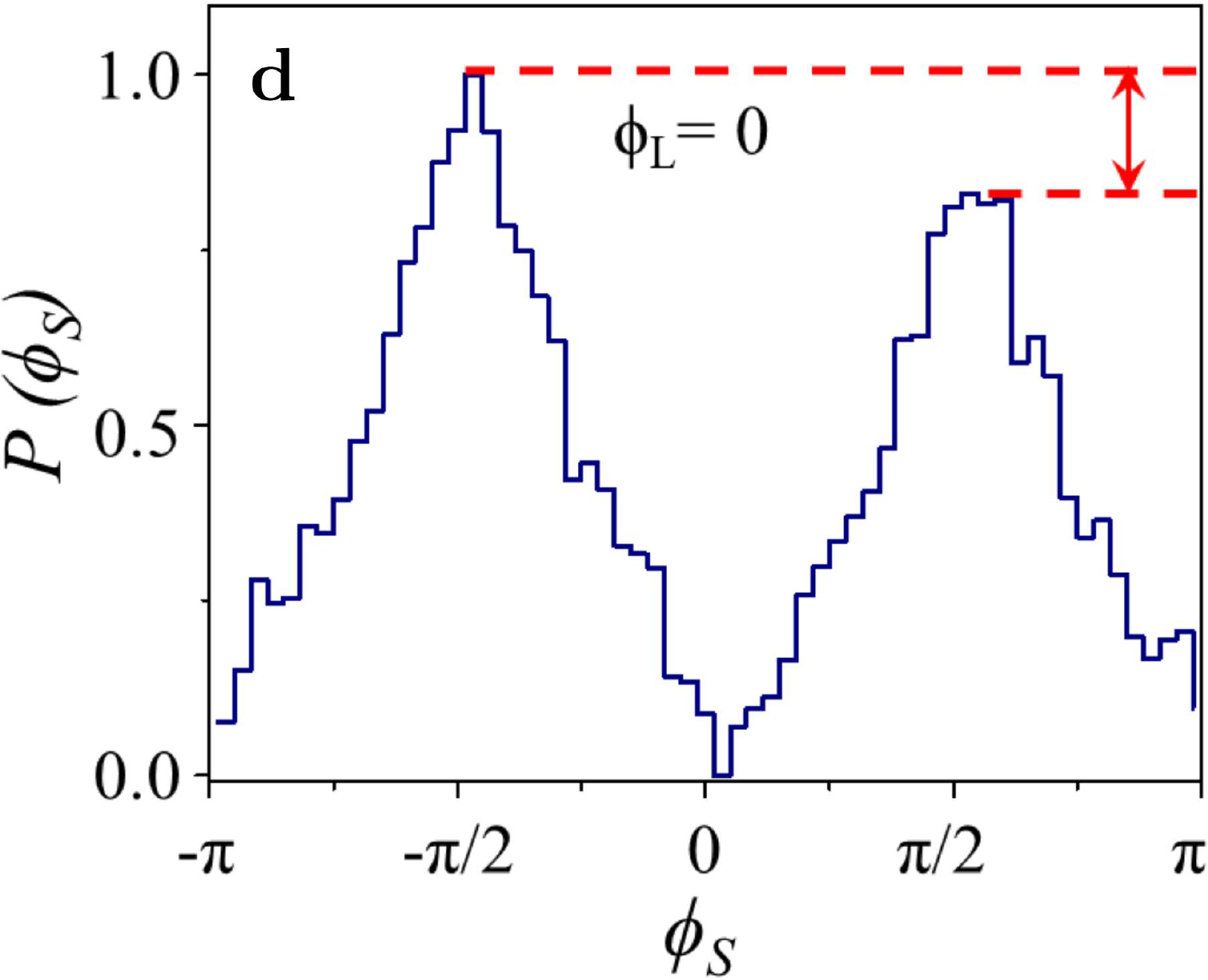}$\qquad$\includegraphics[scale=0.4]{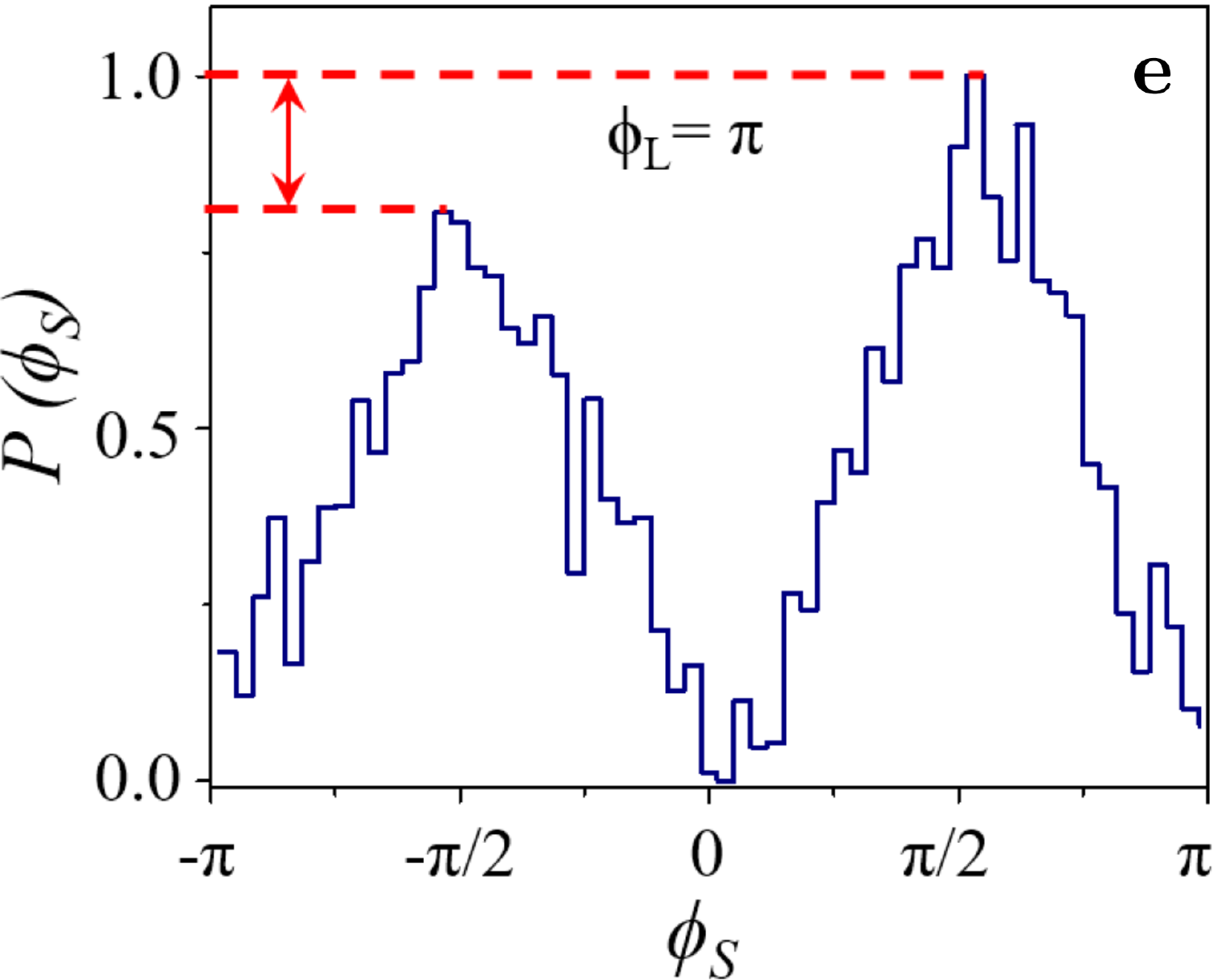}$\quad$
\par\end{centering}
\caption{\textbf{Coincidentally measured momentum distributions of S$^{+}$
and O$^{+}$.} Here $p_{Y}$ and $p_{Z}$ are the projections of fragments'
momenta on $Y$ and $Z$ axes, respectively (measured in atomic units).
\textbf{a}, Isotropic momentum distribution for S$^{+}$ and O$^{+}$
ions measured before the arrival of OTC pulse. \textbf{b} and \textbf{c},
Anisotropic momentum distributions for S$^{+}$ and O$^{+}$ ions
measured at $t\approx0.20\mathrm{\;ps}$ after the application of
the OTC pulse at $\phi_{L}=$ 0 and $\pi$, respectively. \textbf{d}
and \textbf{e}, Angular distributions of S$^{+}$ derived from \textbf{b}
and \textbf{c,} respectively. \label{fig:Figure 4 - Experimental Results}}
\end{figure*}

As seen, the functional form of $U\left(\theta\right)$ is an upward
opening parabola both at $\theta=0$ and $\theta=\pi$. However, the
depths of these parabolas (relative difference is $2B$), as well
as stiffnesses, $\kappa_{\mp}$ ($0<\kappa_{-},\kappa_{+}$) differ.
In addition, the maximum of the potential shifts either to the left
($\phi_{L}<\pi/2$), meaning that more molecules are kicked towards
$\theta=\pi$, or to the right ($\phi_{L}>\pi/2$), in which case
more molecules are kicked towards $\theta=0$. Moreover, the focusing
times at $\theta=0,\pi$ depend on the stiffnesses, $\kappa_{\mp}$.
For example, in the case of $\phi_{L}=0$ (blue-$\vartriangle$ curve,
Fig. \ref{fig:Fig 2 - 1D Model}b), $\kappa_{-}<\kappa_{+}$ and shortly
after the kick, the angular distribution first focuses at $\theta=\pi$,
and afterwards at $\theta=0$, resulting in a pronounced left-right
asymmetry of the angular distribution at the moment of each focusing
event. Being an integrated quantity, the orientation factor, $\braket{\cos\theta}(t)$
is almost insensitive to the sharp features of the distribution. It
does not attain its maximal value at the moment of the highest left-right
asymmetry of the distribution, but rather at the moment when the difference
in the areas under the distribution curve to either side of $\theta=0$
is the largest. For this reason, in discussions of the simulation
and experimental results below we will report both measures of orientation.

The degree of orientation of the S-axis along the $Z$ axis in a molecular
ensemble is determined by the balance between aligning and orienting
interactions with SH field, as described by the first and the second
terms in Eq. \ref{eq:U-afo-theta-special}, respectively. The orientation
effect is emphasized relative to the alignment with reduction of SH
field amplitude, because the aligning interaction is quadratic in
$\mathcal{E}_{20},$ while the orienting interaction is linear. Although
the optimization of the orientation process is not a subject of the
present study, it is worth mentioning that splitting of a two-color
pulse into two subpulses may be beneficial for the overall degree
of orientation \cite{ECNU}.

As a next step, we proceed to a full three-dimensional simulation
of the orientation process. Consider a thermal ($T=18\;K$) ensemble
of $N\gg1$ asymmetric classical rigid rotors with known polarizability
and hyperpolarizability tensors, which are subject to a phase-locked
OTC pulse. To simulate the time-dependent rotational dynamics of the
molecules, we adopt an efficient singularity-free numerical technique,
where quaternions are used to parametrize the rotation \cite{Art-of-Molecular-Simulation,quaternions,Kuipers2002}.
The angular velocity of each molecule is obtained by numerical integration
of the Euler equations \cite{LANDAU}. The orientation in the laboratory
frame of reference is retrieved from numerical integration of the
angular-velocity-dependent equation of motion for the quaternion.
A detailed description of this computational approach can be found
in the Methods Subsection \ref{subsec:methods-2}.

In our Monte Carlo simulations we used ensembles of $N=500,000$ molecules
that are initially isotropically distributed. To account for the essentially
2D character of the detection setup (Coulomb explosion by a probe
pulse circularly polarized in the $YZ$ plane), and to approximate
the experimental measurement conditions, the observable quantities
were calculated by averaging over a sub-ensemble of molecules lying
approximately in the $YZ$ plane at the time of the measurement. The
selection criterion was based on the angles of both minor and major
axes with respect to the $YZ$ plane. Only when both of them were
simultaneously less than $\pi/4$ (a value close to the experimental
arrangement) the molecule was taken into account. The peak intensities
of the pulses were $I_{\mathrm{FW}}=1.4\times10^{14}\ {\rm W/cm}^{2}$
and $I_{\mathrm{SH}}=0.3\times10^{14}\ {\rm W/cm}^{2}$ and the duration
(FWHM) of the pulses was $120\ {\rm fs}$, which is much shorter than
the typical periods of molecular rotation.

Figures \ref{fig:Figure 3 - Results of 3D Simulation}a,b plot the
simulated orientation of the S-axis (characterized by the $\langle\cos\phi_{\mathrm{S}Z}\rangle$)
and alignment of the O-axis (characterized by the $\langle\cos^{2}\phi_{\mathrm{O}}\rangle$)
for various time delays with respect to the OTC pulse ($\phi_{L}=0$).
Here $\phi_{\mathrm{O}}$($\phi_{\mathrm{S}Z}$) are the angles that
the projections of the O-axis(S-axis) on the $YZ$ plane constitute
with respect to the $Y$($Z$) laboratory axis. For an isotropic molecular
ensemble, $\langle\cos^{2}\phi_{\mathrm{O}}\rangle=0.5$ and $\langle\cos\phi_{\mathrm{S}Z}\rangle=0.0$.
For the example shown in Fig. \ref{fig:Figure 3 - Results of 3D Simulation},
both the alignment and the orientation reach maximal values at about
$0.22$ ps after the excitation by the OTC pulse. Three-dimensional
surface plots of the time-dependent probability distributions $P\left(\phi_{\mathrm{\mathrm{S}}},t\right)$
and $P\left(\phi_{\mathrm{\mathrm{O}}},t\right)$ are depicted in
Figs. \ref{fig:Figure 3 - Results of 3D Simulation}c and \ref{fig:Figure 3 - Results of 3D Simulation}d.
Here $\phi_{\mathrm{S}}$ are the angles that projections of the S-axis
on the $YZ$ plane constitute with respect to the $Y$ laboratory
axis. Figure \ref{fig:Figure 3 - Results of 3D Simulation}c shows
the expected \emph{asymmetric} focusing of the angular distribution
of the S-axis at $\phi_{\mathrm{S}}=-\pi/2,\pi/2$ after $0.16$ ps.
In analogy to the 2D model considered above, the focusing at $\phi_{\mathrm{S}}=-\pi/2$
slightly precedes the one at $\phi_{\mathrm{S}}=\pi/2$ (see Fig.
\ref{fig:Figure 3 - Results of 3D Simulation}e). Moreover, Figure
\ref{fig:Figure 3 - Results of 3D Simulation}f shows a pronounced
left-right asymmetry in the distribution of the $\phi_{\mathrm{S}}$
angle at $t=0.16\;\mathrm{ps}$ (ratio of peaks' heights is $\sim0.8$).
The orientation factor attains its maximal value some later ($t=0.22\;\mathrm{ps}$),
when the difference of the areas under the distribution curve to either
side of $\phi_{\mathrm{S}}=0$ is the largest (see Figure \ref{fig:Figure 3 - Results of 3D Simulation}g).
Thus, the effect of orientation is captured by both the assymetry
of the distribution, $P\left(\phi_{\mathrm{\mathrm{S}}},t\right)$
and the integrated quantity, $\langle\cos\phi_{\mathrm{S}Z}\rangle\left(t\right)$.
Figure \ref{fig:Figure 3 - Results of 3D Simulation}d clearly demonstrates
the expected symmetric ($\phi_{\mathrm{O}}=0,\pi$) focusing of angular
distribution of the O-axis, which happens simultaneously with the
orientation of the S-axis.

Figures \ref{fig:Figure 4 - Experimental Results}a-c show the experimentally
measured momentum distributions of the coincidentally measured S$^{+}$
(vertical plane) and O$^{+}$ (horizontal plane) ions ejected from
the Coulomb exploded triply ionized SO$_{2}$ molecules at 0.20 ps
after the application of the OTC pulse. The raw data was normalized
to compensate for the detector bias and possible imperfection in the
circularity of the probe pulse. The details of the data processing
procedure are given in the Methods Subsection \ref{subsec:methods-1}.
For the reference, the momentum distribution of the ions ejected from
molecules exploded before the application of the OTC pulse (i.e. at
negative time delay) is presented in Fig. \ref{fig:Figure 4 - Experimental Results}a.
In this case, an isotropic angular distribution for both axes is clearly
seen. After the application of the OTC pulse, the evident alignment
of the major O-axis along the $Y$ direction (FW polarization) can
be seen on the horizontal plane for both laser phases $\phi_{L}=0$
and $\phi_{L}=\pi$ (Figs. \ref{fig:Figure 4 - Experimental Results}b
and \ref{fig:Figure 4 - Experimental Results}c). For the minor S-axis,
a similar angular clustering along $Z$ axis, i.e. the direction of
the SH polarization, is achieved while exhibiting a noticeable left-right
asymmetric distribution. This asymmetry along the $Z$ axis stands
for the laser-induced orientation of the molecular S-axis, which can
be controlled by adjusting the laser phase $\phi_{L}$ of the OTC
pulse. Figures \ref{fig:Figure 4 - Experimental Results}d and \ref{fig:Figure 4 - Experimental Results}e
plot the corresponding angular distributions of the S-axis for the
phases of $\phi_{L}=$ 0 and $\pi$, respectively. The orientation
degree of the S-axis is estimated to be $\langle\cos\phi_{\mathrm{S}Z}\rangle=-0.069$
for $\phi_{L}=$ $0$ and $\langle\cos\phi_{\mathrm{S}Z}\rangle=0.043$
for $\phi_{L}=\pi$, respectively. These values are in quantitative
agreement with the numerical simulations. The alignment degree of
the O-axis is about $\langle\cos^{2}\phi_{\mathrm{O}}\rangle=0.79$.

So far, the observed data had been discussed in the traditional terms
of the degree of orientation, $\langle\cos\phi_{\mathrm{S}Z}\rangle$.
This is the proper framework when one is dealing with optical experiments,
where experimental observation reflect the integrated index of refraction,
namely a quantity averaged over all possible molecular orientations.
However, COLTRIMS is different and uniquely offers detailed information
on the probability of finding molecules at a specific angle (see Figs.
\ref{fig:Figure 3 - Results of 3D Simulation}c,f for the theoretical
results and Figs. \ref{fig:Figure 4 - Experimental Results}d,e for
the experimental data). As seen in these figures, there is a pronounced
difference in the peak values of the distribution for the left $\phi_{\mathrm{S}}=-\pi/2$
and the right facing molecules $\phi_{\mathrm{S}}=\pi/2$. Therefore,
we introduce an additional measure of orientation, the differential
degree of orientation (DDO), defined as
\begin{equation}
\mathrm{DDO}=\frac{P(\phi_{\mathrm{S}}=\pi/2)-P(\phi_{\mathrm{S}}=-\pi/2)}{P(\phi_{\mathrm{S}}=-\pi/2)+P(\phi_{\mathrm{S}}=\pi/2)}.\label{eq:DDO}
\end{equation}
This measure of orientation is not addressable by the optical detection
schemes, but is readily observable in our COLTRIMS experiments. The
DDO value obtained from the simulation results is $0.10$, while the
experimentally measured ones are $\mathrm{DDO}_{\phi_{L}=0}=-0.09$
and $\mathrm{DDO}_{\phi_{L}=\pi}=0.11$. This indicates the ability
of the OTC pulse to induce relatively high degree of asymmetry in
the otherwise uniform angular distribution of the $\phi_{\mathrm{S}}$
angle.

In summary, while various different combinations of single and multiple
ultrashort pulse excitation were shown to cause molecular alignment,
in order to achieve molecular orientation, symmetry breaking must
be induced. We have demonstrated, experimentally and theoretically,
that field-free three-dimensional molecular orientation can be achieved
by phase-locked OTC laser pulse. The relative phase between the two-color
fields determines the direction of the oriented molecules in space.
The degree of orientation can be controlled by optimizing the parameters
(peak intensity and pulse duration) of the laser fields at the fundamental
and second harmonic frequencies. For a given total available laser
pulse energy, optimal allocation of energy to each component, and/or
splitting an OTC pulse in several subpulses may further increase the
degree of orientation. To the best of our knowledge, our work provides
the first demonstration of field-free, all optical three-dimensional
orientation of asymmetric-top molecules. Our approach may find use
in the eventual orientation of larger molecules as a step towards
their structural analysis by attosecond imaging methodologies \cite{molimaging}.

\section*{References}

\bibliographystyle{unsrt}

\begin{thebibliography}{10}

\bibitem{Velotta2001}
R.~Velotta, N.~Hay, M.~B. Mason, M.~Castillejo, and J.~P. Marangos.
\newblock High-order harmonic generation in aligned molecules.
\newblock {\em Phys. Rev. Lett.}, 87:183901, Oct 2001.

\bibitem{Itatani2005}
J.~Itatani, D.~Zeidler, J.~Levesque, Michael Spanner, D.~M. Villeneuve, and
  P.~B. Corkum.
\newblock Controlling high harmonic generation with molecular wave packets.
\newblock {\em Phys. Rev. Lett.}, 94:123902, Mar 2005.

\bibitem{Patchkovskii2007}
Serguei Patchkovskii, Zengxiu Zhao, Thomas Brabec, and D.~M. Villeneuve.
\newblock High harmonic generation and molecular orbital tomography in
  multielectron systems.
\newblock {\em The Journal of Chemical Physics}, 126(11):114306, 2007.

\bibitem{Lein2007}
Manfred Lein.
\newblock Molecular imaging using recolliding electrons.
\newblock {\em Journal of Physics B: Atomic, Molecular and Optical Physics},
  40(16):R135, 2007.

\bibitem{Larsen1999}
Jakob~Juul Larsen, Ida Wendt-Larsen, and Henrik Stapelfeldt.
\newblock Controlling the branching ratio of photodissociation using aligned
  molecules.
\newblock {\em Phys. Rev. Lett.}, 83:1123--1126, Aug 1999.

\bibitem{Tsubouchi2001}
Masaaki Tsubouchi, Benjamin~J. Whitaker, Li~Wang, Hiroshi Kohguchi, and
  Toshinori Suzuki.
\newblock Photoelectron imaging on time-dependent molecular alignment created
  by a femtosecond laser pulse.
\newblock {\em Phys. Rev. Lett.}, 86:4500--4503, May 2001.

\bibitem{Litvinyuk2003}
I.~V. Litvinyuk, Kevin~F. Lee, P.~W. Dooley, D.~M. Rayner, D.~M. Villeneuve,
  and P.~B. Corkum.
\newblock Alignment-dependent strong field ionization of molecules.
\newblock {\em Phys. Rev. Lett.}, 90:233003, Jun 2003.

\bibitem{Kupper2014}
Jochen K\"upper, Stephan Stern, Lotte Holmegaard, Frank Filsinger, Arnaud
  Rouz\'ee, Artem Rudenko, Per Johnsson, Andrew~V. Martin, Marcus Adolph,
  Andrew Aquila, Sa\ifmmode \check{s}\else~\v{s}\fi{}a Bajt, Anton Barty,
  Christoph Bostedt, John Bozek, Carl Caleman, Ryan Coffee, Nicola Coppola,
  Tjark Delmas, Sascha Epp, Benjamin Erk, Lutz Foucar, Tais Gorkhover, Lars
  Gumprecht, Andreas Hartmann, Robert Hartmann, G\"unter Hauser, Peter Holl,
  Andre H\"omke, Nils Kimmel, Faton Krasniqi, Kai-Uwe K\"uhnel, Jochen Maurer,
  Marc Messerschmidt, Robert Moshammer, Christian Reich, Benedikt Rudek, Robin
  Santra, Ilme Schlichting, Carlo Schmidt, Sebastian Schorb, Joachim Schulz,
  Heike Soltau, John C.~H. Spence, Dmitri Starodub, Lothar Str\"uder, Jan
  Th\o{}gersen, Marc J.~J. Vrakking, Georg Weidenspointner, Thomas~A. White,
  Cornelia Wunderer, Gerard Meijer, Joachim Ullrich, Henrik Stapelfeldt, Daniel
  Rolles, and Henry~N. Chapman.
\newblock X-ray diffraction from isolated and strongly aligned gas-phase
  molecules with a free-electron laser.
\newblock {\em Phys. Rev. Lett.}, 112:083002, Feb 2014.

\bibitem{Glownia2016}
J.~M. Glownia, A.~Natan, J.~P. Cryan, R.~Hartsock, M.~Kozina, M.~P. Minitti,
  S.~Nelson, J.~Robinson, T.~Sato, T.~van Driel, G.~Welch, C.~Weninger, D.~Zhu,
  and P.~H. Bucksbaum.
\newblock Self-referenced coherent diffraction x-ray movie of \aa{}ngstrom- and
  femtosecond-scale atomic motion.
\newblock {\em Phys. Rev. Lett.}, 117:153003, Oct 2016.

\bibitem{Yang2016}
Jie Yang, Markus Guehr, Xiaozhe Shen, Renkai Li, Theodore Vecchione, Ryan
  Coffee, Jeff Corbett, Alan Fry, Nick Hartmann, Carsten Hast, Kareem Hegazy,
  Keith Jobe, Igor Makasyuk, Joseph Robinson, Matthew~S. Robinson, Sharon
  Vetter, Stephen Weathersby, Charles Yoneda, Xijie Wang, and Martin Centurion.
\newblock Diffractive imaging of coherent nuclear motion in isolated molecules.
\newblock {\em Phys. Rev. Lett.}, 117:153002, Oct 2016.

\bibitem{StapelfeldtSeidman2003}
H.~Stapelfeldt and T.~Seideman.
\newblock Colloquium.
\newblock {\em Rev. Mod. Phys.}, 75:543--557, Apr 2003.

\bibitem{Ohshima2010}
Y.~Ohshima and H.~Hasegawa.
\newblock Coherent rotational excitation by intense nonresonant laser fields.
\newblock {\em International Reviews in Physical Chemistry}, 29(4):619--663,
  2010.

\bibitem{Fleischer2012}
S.~Fleischer, Y.~Khodorkovsky, E.~Gershnabel, Y.~Prior, and I.~Sh. Averbukh.
\newblock Molecular alignment induced by ultrashort laser pulses and its impact
  on molecular motion.
\newblock {\em Israel Journal of Chemistry}, 52(5):414--437, 2012.

\bibitem{Lemeshko2013}
M.~Lemeshko, R.~V. Krems, J.~M. Doyle, and S.~Kais.
\newblock Manipulation of molecules with electromagnetic fields.
\newblock {\em Molecular Physics}, 111(12-13):1648--1682, 2013.

\bibitem{molimaging}
Junliang Xu, Cosmin~I Blaga, Pierre Agostini, and Louis~F DiMauro.
\newblock Time-resolved molecular imaging.
\newblock {\em Journal of Physics B: Atomic, Molecular and Optical Physics},
  49(11):112001, 2016.

\bibitem{Friedrich1999}
Bretislav Friedrich and Dudley Herschbach.
\newblock Enhanced orientation of polar molecules by combined electrostatic and
  nonresonant induced dipole forces.
\newblock {\em The Journal of Chemical Physics}, 111(14):6157--6160, 1999.

\bibitem{Sakai2003}
Hirofumi Sakai, Shinichirou Minemoto, Hiroshi Nanjo, Haruka Tanji, and Takayuki
  Suzuki.
\newblock Controlling the orientation of polar molecules with combined
  electrostatic and pulsed, nonresonant laser fields.
\newblock {\em Phys. Rev. Lett.}, 90:083001, Feb 2003.

\bibitem{Ghafur2009}
Omair Ghafur, Arnaud Rouzee, Arjan Gijsbertsen, Wing~Kiu Siu, Steven Stolte,
  and Marc J.~J. Vrakking.
\newblock Impulsive orientation and alignment of quantum-state-selected no
  molecules.
\newblock {\em Nat Phys}, 5(4):289--293, April 2009.

\bibitem{Holmegaard2009}
Lotte Holmegaard, Jens~H. Nielsen, Iftach Nevo, Henrik Stapelfeldt, Frank
  Filsinger, Jochen K\"upper, and Gerard Meijer.
\newblock Laser-induced alignment and orientation of quantum-state-selected
  large molecules.
\newblock {\em Phys. Rev. Lett.}, 102:023001, Jan 2009.

\bibitem{Goban2008}
Akihisa Goban, Shinichirou Minemoto, and Hirofumi Sakai.
\newblock Laser-field-free molecular orientation.
\newblock {\em Phys. Rev. Lett.}, 101:013001, Jun 2008.

\bibitem{Harde1991}
H.~Harde, S\o{}ren Keiding, and D.~Grischkowsky.
\newblock Thz commensurate echoes: Periodic rephasing of molecular transitions
  in free-induction decay.
\newblock {\em Phys. Rev. Lett.}, 66:1834--1837, Apr 1991.

\bibitem{Dion1999}
C.M. Dion, A.D. Bandrauk, O.~Atabek, A.~Keller, H.~Umeda, and Y.~Fujimura.
\newblock Two-frequency ir laser orientation of polar molecules. numerical
  simulations for {HCN}.
\newblock {\em Chemical Physics Letters}, 302(3-4):215--223, 1999.

\bibitem{Averbukh2001}
I.~Sh. Averbukh and R.~Arvieu.
\newblock Angular focusing, squeezing, and rainbow formation in a strongly
  driven quantum rotor.
\newblock {\em Phys. Rev. Lett.}, 87:163601, Sep 2001.

\bibitem{Machholm2001}
Mette Machholm and Niels~E. Henriksen.
\newblock Field-free orientation of molecules.
\newblock {\em Phys. Rev. Lett.}, 87:193001, Oct 2001.

\bibitem{Fleischer2011}
Sharly Fleischer, Yan Zhou, Robert~W. Field, and Keith~A. Nelson.
\newblock Molecular orientation and alignment by intense single-cycle thz
  pulses.
\newblock {\em Phys. Rev. Lett.}, 107:163603, Oct 2011.

\bibitem{Kitano2013}
Kenta Kitano, Nobuhisa Ishii, Natsuki Kanda, Yoshiyuki Matsumoto, Teruto Kanai,
  Makoto Kuwata-Gonokami, and Jiro Itatani.
\newblock Orientation of jet-cooled polar molecules with an intense
  single-cycle thz pulse.
\newblock {\em Phys. Rev. A}, 88:061405, Dec 2013.

\bibitem{Daems2005}
D.~Daems, S.~Gu\'erin, D.~Sugny, and H.~R. Jauslin.
\newblock Efficient and long-lived field-free orientation of molecules by a
  single hybrid short pulse.
\newblock {\em Phys. Rev. Lett.}, 94:153003, Apr 2005.

\bibitem{Gershnabel2006}
E.~Gershnabel, I.~Sh. Averbukh, and Robert~J. Gordon.
\newblock Orientation of molecules via laser-induced antialignment.
\newblock {\em Phys. Rev. A}, 73:061401, Jun 2006.

\bibitem{Egodapitiya2014}
K.~N. Egodapitiya, Sha Li, and R.~R. Jones.
\newblock Terahertz-induced field-free orientation of rotationally excited
  molecules.
\newblock {\em Phys. Rev. Lett.}, 112:103002, Mar 2014.

\bibitem{Damari2016}
Ran Damari, Shimshon Kallush, and Sharly Fleischer.
\newblock Rotational control of asymmetric molecules: Dipole- versus
  polarizability-driven rotational dynamics.
\newblock {\em Phys. Rev. Lett.}, 117:103001, Sep 2016.

\bibitem{Gershnabel2018}
E.~Gershnabel and I.~Sh. Averbukh.
\newblock Orienting asymmetric molecules by laser fields with twisted
  polarization.
\newblock {\em Phys. Rev. Lett.}, 120:083204, Feb 2018.

\bibitem{Yachmenev2016}
A.~Yachmenev and S.~N. Yurchenko.
\newblock Detecting chirality in molecules by linearly polarized laser fields.
\newblock {\em Phys. Rev. Lett.}, 117:033001, Jul 2016.

\bibitem{Tutunnikov2018}
Ilia Tutunnikov, Erez Gershnabel, Shachar Gold, and Ilya~Sh. Averbukh.
\newblock Selective orientation of chiral molecules by laser fields with
  twisted polarization.
\newblock {\em The Journal of Physical Chemistry Letters}, 9(5):1105--1111,
  2018.
\newblock PMID: 29417812.

\bibitem{Vrakking1997}
Marc.~J.J. Vrakking and Steven Stolte.
\newblock Coherent control of molecular orientation.
\newblock {\em Chemical Physics Letters}, 271(4-6):209--215, 1997.

\bibitem{Kanai2001}
Tsuneto Kanai and Hirofumi Sakai.
\newblock Numerical simulations of molecular orientation using strong,
  nonresonant, two-color laser fields.
\newblock {\em The Journal of Chemical Physics}, 115(12):5492--5497, 2001.

\bibitem{De2009}
S.~De, I.~Znakovskaya, D.~Ray, F.~Anis, Nora~G. Johnson, I.~A. Bocharova,
  M.~Magrakvelidze, B.~D. Esry, C.~L. Cocke, I.~V. Litvinyuk, and M.~F. Kling.
\newblock Field-free orientation of co molecules by femtosecond two-color laser
  fields.
\newblock {\em Phys. Rev. Lett.}, 103:153002, Oct 2009.

\bibitem{Oda2010}
Keita Oda, Masafumi Hita, Shinichirou Minemoto, and Hirofumi Sakai.
\newblock All-optical molecular orientation.
\newblock {\em Phys. Rev. Lett.}, 104:213901, May 2010.

\bibitem{JW2010}
Jian Wu and Heping Zeng.
\newblock Field-free molecular orientation control by two ultrashort dual-color
  laser pulses.
\newblock {\em Phys. Rev. A}, 81:053401, May 2010.

\bibitem{Frumker2012}
E.~Frumker, C.~T. Hebeisen, N.~Kajumba, J.~B. Bertrand, H.~J. W\"orner,
  M.~Spanner, D.~M. Villeneuve, A.~Naumov, and P.~B. Corkum.
\newblock Oriented rotational wave-packet dynamics studies via high harmonic
  generation.
\newblock {\em Phys. Rev. Lett.}, 109:113901, Sep 2012.

\bibitem{Takemoto2008}
Norio Takemoto and Kaoru Yamanouchi.
\newblock Fixing chiral molecules in space by intense two-color phase-locked
  laser fields.
\newblock {\em Chemical Physics Letters}, 451(1):1 -- 7, 2008.

\bibitem{DORNER200095}
R.~D\"{o}rner, V.~Mergel, O.~Jagutzki, L.~Spielberger, J.~Ullrich,
  R.~Moshammer, and H.~Schmidt-B\"{o}cking.
\newblock Cold target recoil ion momentum spectroscopy: a momentum microscope
  to view atomic collision dynamics.
\newblock {\em Physics Reports}, 330(2):95--192, 2000.

\bibitem{MolecularHyperpolarizabilities}
A.~D. Buckingham and B.~J. Orr.
\newblock Molecular hyperpolarisabilities.
\newblock {\em Q. Rev. Chem. Soc.}, 21:195--212, 1967.

\bibitem{SO2-Data}
George Maroulis.
\newblock The electric hyperpolarizability of ozone and sulfur dioxide.
\newblock {\em Chemical Physics Letters}, 189(2):112--118, 1992.

\bibitem{ECNU}
Shian Zhang, Chenhui Lu, Tianqing Jia, Zugeng Wang, and Zhenrong Sun.
\newblock Field-free molecular orientation enhanced by two dual-color laser
  subpulses.
\newblock {\em The Journal of Chemical Physics}, 135:034301, 2011.

\bibitem{Art-of-Molecular-Simulation}
D.~C. Rapaport.
\newblock {\em The Art of Molecular Dynamics Simulation}.
\newblock Cambridge University Press, 2 edition, 2004.

\bibitem{quaternions}
L.~Romero E.~A.~Coutsias.
\newblock The quaternions with an application to rigid body dynamics.
\newblock {T}echreport, University of New Mexico, Albuquerque, NM 87131, Feb
  1999.

\bibitem{Kuipers2002}
J.~B. Kuipers.
\newblock {\em Quaternions and Rotation Sequences: A Primer with Applications
  to Orbits, Aerospace and Virtual Reality}.
\newblock Princeton University Press, 2002.

\bibitem{LANDAU}
L.D. Landau and E.M. Lifshitz.
\newblock {\em Mechanics}.
\newblock Butterworth-Heinemann, Oxford, {T}hird edition, 1976.

\bibitem{SamplingSO3}
S.~M. LaValle.
\newblock {\em Planning Algorithms}.
\newblock Cambridge University Press, New York, NY, USA, 2006.

\end{thebibliography}

\section*{Acknowledgments}

This work is supported by the National Natural Science Fundation of
China (Grant Nos. 11425416, 61690224 and 11761141004), the 111 Project
of China (Grant No. B12024), the Israel Science Foundation (Grant
No. 746/15), the ICORE program ``Circle of Light\textquotedblright ,
and the ISF-NSFC (Grant No. 2520/17). I.A. acknowledges support as
the Patricia Elman Bildner Professorial Chair. This research was made
possible in part by the historic generosity of the Harold Perlman
Family.

\section*{Author contributions}

K.L. and I. T. contributed equally to this work. K.L., J.Q., J.M.,
Q.S., Q.J., W.Z., H.L., F.S., X.G., H.L., P.L., H.Z. and J.W. designed
and performed the experiments, I.T. and I.A. carried out the theoretical
analysis and numerical simulations. Y.P., I.A and J.W. jointly supervised
the whole project. K.L., I.T., Y.P., I.A., and J.W. wrote the manuscript.
All authors contributed to the analyses and discussions of the results.

\section*{Competing financial interests}

The authors declare no competing financial interests.

\section*{Methods}

\subsection{Experimental Methods \label{subsec:methods-1}}

Field-free 3D orientation of $\mathrm{SO}_{2}$ molecule in a molecular
beam is induced by a pair of orthogonally polarized two-color femtosecond
laser pulses. Following the orientation, a intense circularly polarized
probe pulse Coulomb-explodes the molecules to image their spatial
orientation at various time delays, as schematically shown in Fig.
\ref{fig:Fig 1 - Experimental Setup}. The output ($25\;\mathrm{fs}$,
$790\;\mathrm{nm}$, $10\;\mathrm{kHz}$) of a Femtolasers multipass
amplifier Ti:sapphire laser system is split into pump and probe arms
via a beam splitter of $7:3$ intensity ratio. The OTC pulse is generated
in a colinear scheme by down-collimating the $70\%$ pump beam into
a $150\;\mathrm{\lyxmathsym{\textmu}m}$-thick $\beta$-barium borate
($\beta$-BBO) crystal to generate a SH pulse at $395\;\mathrm{nm}$.
To increase the doubling efficiency, a telescope is placed in front
of the $\beta$-BBO crystal to reduce the beam diameter by a factor
of two. A path of $7\;\mathrm{mm}$-thick $\alpha$-barium borate
($\alpha$-BBO) crystals is introduced after the $\beta$-BBO crystal
to compensate the group delay between the two colors that is induced
by the optical components (wedges, mirrors and windows) along the
beams' path. The $30\%$ probe beam is passed through a quarter wave
plate, followed by a beam expander, making it circularly polarized
and doubles its diameter. A motorized delay stage in the probe arm
is used to synchronize and adjust its time delay with respect to the
OTC pulse. The two pulses are afterwards focused onto a supersonic
molecular beam of $20\%$ mixture of $\mathrm{SO}_{2}$ in He in an
ultrahigh vacuum chamber of the COLTRIMS apparatus by a concave silver
mirror ($f=7.5\;\mathrm{cm}$).

By pre-compensation the pulse chirp prior to the amplifier, the temporal
duration of the probe pulse in the interaction region is controlled
to be $\sim40\;\mathrm{fs}$. The OTC pulse is stretched to be $\sim120\;\mathrm{fs}$
after the BBO crystals, wedge pair, the combination mirror and the
entrance window. The intensities of the FW and the SH in the reaction
area are measured to be $I_{\mathrm{FH}}\approx1.4\times10^{14}\;\mathrm{W/cm^{2}}$,
$I_{\mathrm{SH}}\approx0.3\times10^{14}\;\mathrm{W/cm^{2}}$ and the
intensity of the probe pulse is $\sim6\times10^{14}\;\mathrm{W/cm^{2}}$.
The rotational temperature of the molecular beam is close to the translation
temperature, which can be estimated from $T_{\mathrm{trans}}=\Delta p^{2}/\left[4\ln\left(4\right)k_{B}m\right]$,
where $k_{B}$ is the Boltzmann's constant, $\Delta p$ and $m$ are
the full-width at half-maximum of the momentum distribution (in the
jet direction) and mass of the singly ionized $\mathrm{SO}_{2}^{+}$,
respectively. In our experiment we measure a momentum width in the
jet direction of $\Delta p\sim6.1\;\mathrm{a.u.}$ of $\mathrm{SO}_{2}^{+}$
ions created by a laser field polarized along the $Z$ axis (orthogonal
to the jet direction). The rotational temperature of the $\mathrm{SO}_{2}$
molecule is estimated to be $18\;K$. The produced fragment ions are
accelerated and guided by a weak homogeneous static electric field
($\sim20\;\mathrm{V/cm}$) and then detected by a time- and position-sensitive
microchannel plate detector. The three-dimensional momenta of the
ions are retrieved from the measured time-of-flights and positions
of the impacts. Here, for the asymmetric-top molecules, the direction
of the principle axes is retrieved from the coincidentally measured
fragment ions of the triple-ionization-induced Coulomb explosion channel
of $\left[\mathrm{SO_{2}}+n\lyxmathsym{\textcrh\textcloseomega}\rightarrow\mathrm{S}^{+}+\mathrm{O}^{+}+\mathrm{O}^{+}+3e\right]$
. The angular distributions for $\phi_{\mathrm{S}}$ and $\phi_{\mathrm{O}}$
away from the $Y$ axis at maximum 3D orientation are measured by
fixing the delay stage around $0.20\;\mathrm{ps}$. To increase the
visibility and eliminate the bias induced by the imperfect circularity
of the probe pulse, the angular distribution at negative time delay
is collected as reference for the data analysis. We normalize the
total probability of the angular distribution to unity for each time
delay and then subtract the averaged angular distribution at negative
times. Since the fragmentation of triply ionized $\mathrm{SO}_{2}$
molecule happens mostly in the polarization plane of the probe pulse,
the data analysis are restricted to this plane by selecting molecules
confined to $\left[-\pi/4,\pi/4\right]$ with respect to the $YZ$
plane.

\subsection{Numerical Methods \label{subsec:methods-2}}

\begin{figure}[h]
\begin{centering}
\includegraphics[scale=0.18]{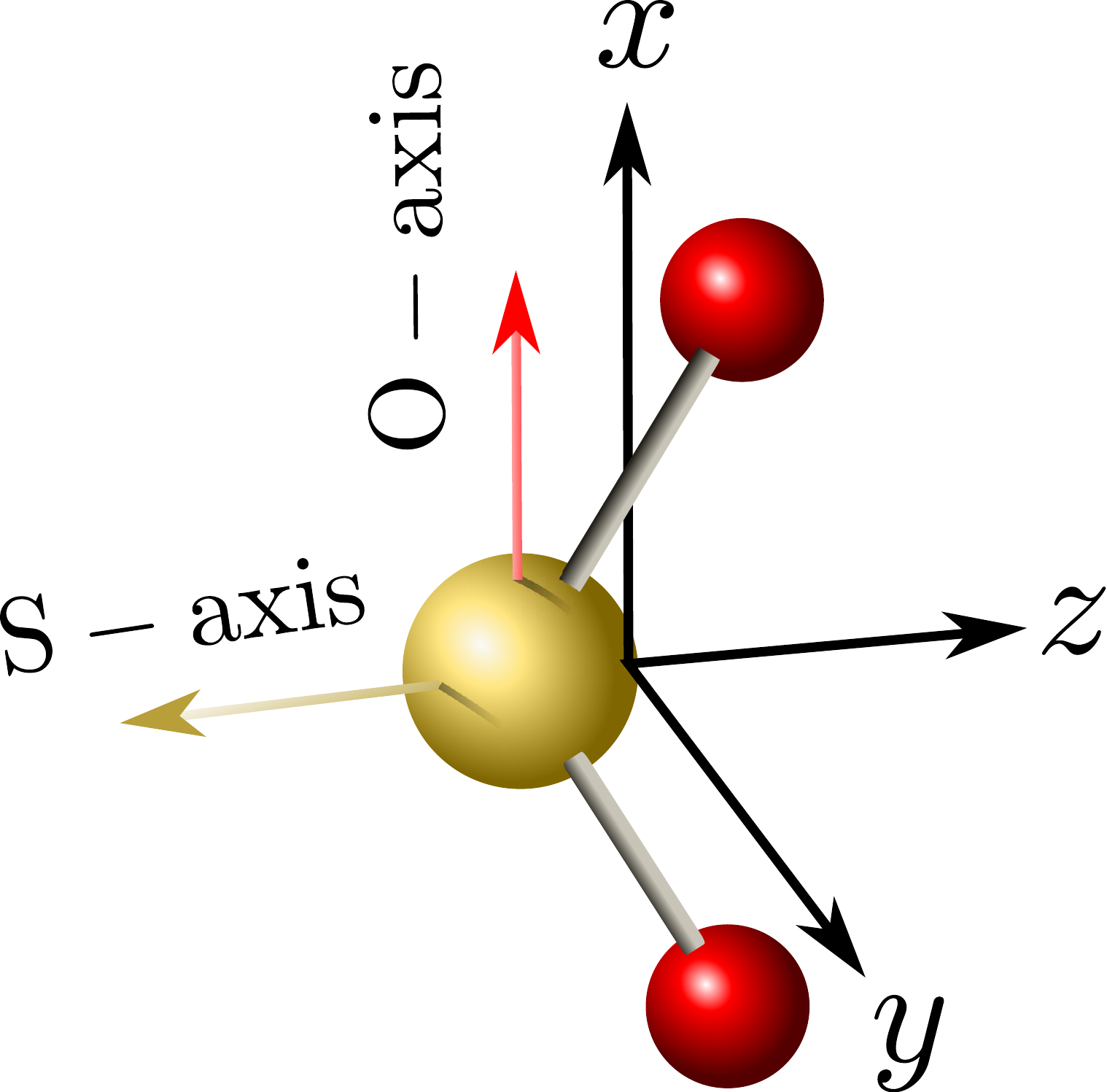}
\par\end{centering}
\caption{The $\mathrm{SO}_{2}$ molecule. Axes $x$, $y$ and $y$ are the
principal axes of the molecule. Here, all atoms lie in $xz$ plane
and color-coded: red - oxygen, yellow - sulfur.\label{fig:the-molecule}}
\end{figure}

We consider the asymmetric molecules as classical rigid rotors with
anisotropic polarizability and hyperpolarizability. A specific example
of the $\mathrm{SO}_{2}$ molecule is presented in Figure \ref{fig:the-molecule}.

Table \ref{tab:table-1} summarizes the properties of the $\mathrm{SO}_{2}$
molecule. The moment of inertia tensor and its principal axes were
computed based on the Cartesian atomic coordinates (measured in bohr)
\cite{SO2-Data}: $\mathrm{S}=\left(0.0,0.0,-0.682958\right)$ and
$\mathrm{O}=\left(\pm2.333567,0.0,0.682581\right)$.
\begin{table}[H]
\begin{centering}
\begin{tabular}{>{\raggedright}p{1.8cm}|>{\raggedright}p{1.8cm}|>{\raggedright}p{2cm}|>{\raggedright}p{1.8cm}}
{\footnotesize{}$\overset{\text{\text{\tiny\ensuremath{\bm{\leftrightarrow}}}}}{\mathbf{I}}$
comp.}  & {\footnotesize{}$\overset{\text{\text{\tiny\ensuremath{\bm{\leftrightarrow}}}}}{\boldsymbol{\alpha}}$comp.}  & {\footnotesize{}$\overset{\text{\text{\tiny\ensuremath{\bm{\leftrightarrow}}}}}{\boldsymbol{\beta}}$
com.}  & {\footnotesize{}$\mathbf{D}$ comp.}\tabularnewline
\hline
{\footnotesize{}$I_{x}=55509$}  & {\footnotesize{}$\alpha_{xx}=31.26$}  & {\footnotesize{}$\beta_{xxz}=22.0$}  & {\footnotesize{}$\mu_{x}=0.0$}\tabularnewline
{\footnotesize{}$I_{y}=371885$}  & {\footnotesize{}$\alpha_{yy}=18.64$}  & {\footnotesize{}$\beta_{yyz}=26.5$}  & {\footnotesize{}$\mu_{y}=0.0$}\tabularnewline
{\footnotesize{}$I_{z}=317477$}  & {\footnotesize{}$\alpha_{zz}=20.80$}  & {\footnotesize{}$\beta_{zzz}=6.4$}  & {\footnotesize{}$\mu_{z}=-0.79$}\tabularnewline
\end{tabular}
\par\end{centering}
\caption{Summary of the $\mathrm{SO}_{2}$ properties (measured in a.u): eigenvalues
of the moment of inertia tensor, components of polarizability tensor,
components of hyperpolarizability tensor and components of dipole
moment in the body-fixed frame of molecular principal axes. \label{tab:table-1}}
\end{table}

We investigate the behavior of an ensemble of $N\gg1$ molecules with
the help of the Monte Carlo simulation in which rotational dynamics
of each molecule is treated numerically. The description of rotational
dynamics in terms of Euler angles is known to lead to singular equations
of motion \cite{LANDAU}. Here, we rely on an efficient singularity-free
numerical technique where quaternions are used to parametrize the
rotation \cite{Art-of-Molecular-Simulation,quaternions,Kuipers2002}.
This approach solves the singularity problem and avoids time consuming
calculations of trigonometric functions. The orientation of a rigid
body is described by a quaternion:
\[
q=\left(q_{0},q_{1},q_{2},q_{3}\right)=\left(\cos\frac{\theta}{2},\sin\frac{\theta}{2}\mathbf{p}\right),
\]
where $\mathbf{p}$ is a unit vector defining the direction of rotation
and $\theta$ is the angle of rotation about it. The rate of change
in time of a quaternion is given by
\begin{equation}
\dot{q}=\frac{1}{2}q\Omega,\label{eq:Quaternion-Equation-of-Motion}
\end{equation}
where $\Omega=\left(0,\boldsymbol{\Omega}\right)$ is a pure quaternion
\cite{Kuipers2002,quaternions} constructed from angular velocity
of the molecule, expressed with respect to the body-fixed frame of
molecular principal axes $a$, $b$ and $c$, $\boldsymbol{\Omega}=\left(\Omega_{a},\Omega_{b},\Omega_{c}\right)$.
In Eq. \ref{eq:Quaternion-Equation-of-Motion}, the quaternions multiplication
rule is implied \cite{Kuipers2002,quaternions}. According to Euler
equation \cite{LANDAU}, the rate of change of the angular velocity
expressed with respect to the body-fixed frame is
\begin{equation}
\overset{\text{\text{\tiny\ensuremath{\bm{\leftrightarrow}}}}}{\mathbf{I}}\dot{\boldsymbol{\Omega}}=\left(\overset{\text{\text{\tiny\ensuremath{\bm{\leftrightarrow}}}}}{\mathbf{I}}\boldsymbol{\Omega}\right)\times\boldsymbol{\Omega}+\boldsymbol{\mathrm{T}},\label{eq:Euler-Equation}
\end{equation}
where $\overset{\text{\text{\tiny\ensuremath{\bm{\leftrightarrow}}}}}{\mathbf{I}}$
is the moment of inertia tensor and $\boldsymbol{\mathrm{T}}=\left(\mathrm{T}_{a},\mathrm{T}_{b},\mathrm{T}_{c}\right)$
is the torque, both expressed with respect to the body-fixed frame.
To model the torque due to interaction with an electric field, we
transform the electric field $\boldsymbol{\mathcal{E}}$, expressed
with respect to the laboratory frame of reference, into the body-fixed
frame. The transformation rule is $E=q^{c}\mathcal{E}q$, where $E=\left(0,\mathbf{E}\right)$
and $\mathcal{E}=\left(0,\boldsymbol{\mathcal{E}}\right)$ are pure
quaternions, constructed from the electric field in the body-fixed
frame and $\boldsymbol{\mathcal{E}}$, respectively. A conjugate of
a quaternion $q$ is denoted by $q^{c}$.\cite{Kuipers2002,quaternions}
Again, quaternions multiplication rule is used in the transformation.
The linear polarizability part of the induced dipole moment in the
body-fixed principal axes frame is given by $\mathbf{D}=\overset{\text{\text{\tiny\ensuremath{\bm{\leftrightarrow}}}}}{\boldsymbol{\alpha}}\mathbf{E}$,
where $\overset{\text{\text{\tiny\ensuremath{\bm{\leftrightarrow}}}}}{\boldsymbol{\alpha}}$
is the polarizability tensor in the body-fixed frame. The torque due
to the polarizability is $\boldsymbol{\mathrm{T}}^{\alpha}=\braket{\mathbf{D}\times\mathbf{E}}$,
where $\braket{\cdot}$ is a time average over fast oscillations of
the optical field. The torque due to the hyperpolarizability in the
body fixed frame is given by $\mathrm{T}_{i}^{\beta}=\braket{\varepsilon_{ijk}\beta_{jnm}E_{n}E_{m}E_{k}},$
where $\varepsilon_{ijk}$ is the Levi-Civita symbol. Explicit torques
expressions for the two-color field are
\[
\mathbf{T}^{\alpha}=\frac{\mathcal{E}_{1}^{2}}{2}\left(\overset{\text{\text{\tiny\ensuremath{\bm{\leftrightarrow}}}}}{\boldsymbol{\alpha}}\mathbf{E}_{1}\right)\times\mathbf{E}_{1}+\frac{\mathcal{E}_{2}^{2}}{2}\left(\overset{\text{\text{\tiny\ensuremath{\bm{\leftrightarrow}}}}}{\boldsymbol{\alpha}}\mathbf{E}_{2}\right)\times\mathbf{E}_{2}
\]
and
\[
T_{i}^{\beta}=\frac{\mathcal{E}_{1}^{2}\mathcal{E}_{2}}{4}\varepsilon_{ijk}\left[\beta_{mnj}E_{1m}E_{2n}E_{1k}+\frac{1}{2}\beta_{mnj}E_{1m}E_{1n}E_{2k}\right],
\]
where $\mathbf{E}_{1}$, $\mathbf{E}_{2}$ are the transformed (into
the body-fixed frame) polarization vectors of the FW and SH, respectively.

In our simulations, the initial ensemble of molecules is generated
via a Monte Carlo procedure. The initial orientations of the molecules
and the corresponding quaternions for an isotropic ensemble are generated
by a random uniform sampling over the space of rotations \cite{SamplingSO3}.
We assume that the molecular ensemble is initially at thermal conditions,
and that the molecular angular velocities are distributed according
to:
\[
f\left(\boldsymbol{\Omega}\right)\propto\exp\left[-\frac{\boldsymbol{\Omega}^{T}\overset{\text{\text{\tiny\ensuremath{\bm{\leftrightarrow}}}}}{\mathbf{I}}\boldsymbol{\Omega}}{2k_{B}T}\right]=\prod_{i}\exp\left[-\frac{I_{i}\Omega_{i}^{2}}{2k_{B}T}\right],
\]
where $i=x,y,z$. Since the kinetic energy is a scalar, it is coordinate
invariant so we express it with respect to the body-fixed frame. Here,
$T$ is the temperature of the gas and $k_{B}$ is the Boltzmann constant.
We integrate the system of equations \ref{eq:Quaternion-Equation-of-Motion}
and \ref{eq:Euler-Equation} using the Runge-Kutta integration method.
At each time step, the quaternions are renormalized in order to preserve
the unit norm \cite{Art-of-Molecular-Simulation,Kuipers2002}.
\end{document}